\documentstyle [12pt]{article}
\let\tilde=\widetilde

\def\tr{{\rm tr}}

\def\one#1{#1^{\raise5pt\hbox{$\scriptstyle\!\!\!\!1$}}\,{}}
\def\two#1{#1^{\raise5pt\hbox{$\scriptstyle\!\!\!\!2$}}\,{}}
\def\three#1{#1^{\raise5pt\hbox{$\scriptstyle\!\!\!\!3$}}\,{}}

\def\phi{\varphi}
\def\eps{\varepsilon}
\def\a{\alpha}
\def\b{\beta}
\def\d{\delta}

\def\l{\lambda}

\def\TT{\tilde{T}}
\def\T{{\cal T}}

\def\beq{\begin{equation}}
\def\eeq{\end{equation}}
\def\be{\begin{displaymath}}
\def\ee{\end{displaymath}}
\def\bds{\begin{description}}
\def\eds{\end{description}}
\def\half{\frac{1}{2}}

\newtheorem{th}{Theorem}
\newtheorem{pr}{Proposition}
\newtheorem{guess}{Conjecture}
\def\N{N}
\begin{document}
\title{Separation of Variables in the Classical\\
        Integrable $SL(3)$ Magnetic Chain}
\author{E. K. Sklyanin \thanks{On leave from Steklov Mathematical Institute,
 Fontanka 27, St.Petersburg 191011, Russia.}\\
 RIMS, Kyoto University, Kyoto 606, Japan \\
           and \\
 Research Institute for Theoretical Physics
\thanks{Supported by the Academy of Finland}\\
  Helsinki University, Helsinki 00170, Finland
        }
\date{March 28, 1992}
\maketitle
\vskip-9.5cm
\hskip9cm
\sf RIMS-871; hep-th/9211126 \rm
\vskip9.5cm

{\small {\bf Abstract.} There are two fundamental problems studied by the
theory of hamiltonian integrable systems: integration of equations of motion,
and construction of action-angle variables. The third problem, however,
should be added to the list: separation of variables. Though much
simpler than two others, it has important relations to the quantum
integrability. Separation of variables is constructed for the $SL(3)$ magnetic
 chain --- an
example of integrable model associated to a nonhyperelliptic algebraic curve.
}
\section{Introduction}
   Consider a completely integrable Hamiltonian system with $D$
degrees of freedom. According to
the definition of complete integrability due to Liouville-Arnold \cite{Arn:89}
it means that the  system  possesses
exactly $D$ independent  Hamiltonians $H_j$
commuting with respect to the Poisson bracket
\beq
\label{eq:Hn}
 \{H_j,H_k\}=0\qquad j,k=1,\ldots,D
\eeq

There are three fundamental problems discussed in the theory of integrable
systems. They are listed below in the order of decreasing complexity:
\begin{itemize}
\item Construction of action-angle variables.
\item Integration of equations of motion.
\item Separation of variables.
\end{itemize}

  For the wide class of finite-dimensional integrable systems subject to the
Inverse Spectral Transform Method
an effective integration of equations of motion
can be performed
using the techniques of algebraic geometry \cite{Dub:80}. As for the
effective
construction of the action-angle variables, it is more difficult problem
\cite{FM:76}, especially when the reality conditions are carefully taken into
account \cite{NV:84}. The case of systems associated to hyperelliptic
spectral curves
is studied in detail \cite{FM:76,NV:84}, there being only some preliminary
results for the non-hyperelliptic case \cite{Dik:81}.

   In the present paper the third mentioned problem (separation of variables)
is studied. To be precise,
the separation of variables is understood  here as construction of $D$ pairs
of canonical variables $x_j$, $p_j$ ($j=1,\ldots, D$)
\beq
\label{eq:px}
 \{x_j,x_k\}=\{p_j,p_k\}=0\qquad  \{p_j,x_k\}=\d_{jk}
\eeq
and $D$ functions $\Phi_j$ such that
\beq
\Phi_j(x_j,p_j,H_1,H_2,\ldots,H_D)=0\qquad j=1,2,\ldots,D
\label{eq:sepvar}
\eeq
where $H_j$ are the Hamiltonians (\ref{eq:Hn}) in involution.

 The above definition is a paraphrase of the usual definition of separability
of variables in the Hamilton-Jacobi equation \cite{Arn:89}. Note that the
canonical transformation from the original variables to $(x_j,p_j)$
may not necessarily be a pure coordinate change, as in textbooks on
classical mechanics, but can involve both coordinates and momenta.

   The problem in question, being the simplest of the three,
is rather neglected in the literature on the subject, though,
in our opinion, it deserves  attention at least for two reasons.
First, the variables $(x_j,p_j)$ serve usually as a raw material for
constructing action-angle variables and integrating equations of motion.
Second,
the problem is interesting for the theory of quantum integrability, since
the construction of separated variables usually has direct counterparts in
the quantum case \cite{Skl:30,Skl:32}.

   The construction of the variables $(x_j,p_j)$ is well known for the case
of the hyperelliptic spectral curve \cite{FM:76,NV:84,Skl:14,Skl:XYZ} though
its relation to the separation of variables is not always stated manifestly.
The coordinates $x_j$
are defined as the zeroes of the corresponding
Baker-Akhiezer function, and
the canonically conjugated momenta
$p_j$, or sometimes $\exp p_j$, usually turn out to be eigenvalues of the
corresponding $L$-operator
taken at the values of the spectral parameter equal to $x_j$. The functions
$\Phi_j$ are then simply the characteristic polynomials of the $L$-operator.

%

   In the present paper we study the problem for the nonhyperelliptic case.
The $SL(3)$ classical magnetic chain is chosen as a sample toy-model.
Having in mind subsequent application to the quantum integrability we
make extensive use of the classical $r$-matrix formalism \cite{CISM}.
We consider
complexified version of the model in order to avoid the additional
complications of the real case. Our construction of separated variables
is quite elementary and does not involve any sophisticated
algebro-geometric techniques.

\section{Description of the model}

    The model we are going to describe is the nonhomogeneous classical $SL(\N)$
magnetic chain. It is in a sense generic for the models related to the
$SL(\N)$-invariant classical $r$-matrix \cite{CISM}. For $\N=2$ the model
was introduced in \cite{Ish:82,Pap:87}. The continuous version of the model
was studied earlier in \cite{Takh:PLA,Orph:80,SR:82}.
For a degenerate case (Gaudin model) see Section 5. The quantum version
of the model is well studied by means of the Bethe ansatz method
\cite{Suth:PRB,KR:LOMI,KR:JPA}.

    The model in question is described in terms of the variables
$S_{\a\b}^{(m)}$, $(\a,\b=1,\ldots,\N$; $m=1,\ldots,M$;
$\sum_{\a=1}^\N S_{\a\a}^{(m)}=0)$ subject to the Poisson brackets
\beq
 \{S^{(m)}_{\a_1\b_1},S^{(n)}_{a_2\b_2}\}=
    (S^{(m)}_{a_1\b_2}\d_{a_2\b_1}-S^{(m)}_{a_2\b_1}\d_{a_1\b_2})\d_{mn}
\label{eq:pbS}
\eeq
which define the Kirillov-Kostant Poissonian structure on the direct product
of $M$ orbits of the coadjoint action of the Lie group $SL(\N)$ on $sl(\N)^*$,
see e.g. \cite{CISM}.
It is well known
that the center of the Poisson algebra is generated by the eigenvalues
$l_\a^{(m)}$ of the matrices $S^{(m)}$
\beq
\label{eq:specS}
\det(u+S^{(m)})=\prod_{\a=1}^{\N}(u+l^{(m)}_\a),
    \qquad \sum_{\a=1}^\N l_\a^{(m)}=0
\eeq

   We shall assume that $l_\a^{(m)}$ are fixed numbers. The Poisson bracket
(\ref{eq:pbS}) is thus nondegenerate on the manifold (\ref{eq:specS}) having
dimension $2D=M\N(\N-1)$ for the case of generic orbit (all eigenvalues
of $S^{(m)}$ are distinct). In what follows we always assume that the orbit is
generic.

   Let $Z$ be an invertible $\N\times\N$ number matrix having $\N$ distinct
eigenvalues, let
$\d_m$ $(m=1,\ldots,M)$ be some fixed numbers, and $u$ be a complex
parameter (spectral parameter). Consider the product (monodromy matrix)
\beq
T(u)=Z(u-\d_M+S^{(M)})\ldots(u-\d_2+S^{(2)})(u-\d_1+S^{(1)})
\label{eq:defT}
\eeq

\begin{pr}
Matrix elements of $T(u)$ have the following quadratic Poisson
brackets
\beq
 \{T_{\a_1\b_1}(u),T_{\a_2\b_2}(v)\}=\frac{1}{u-v}
   \bigl(T_{\a_2\b_1}(u)T_{\a_1\b_2}(v)-T_{\a_1\b_2}(u)T_{\a_2\b_1}(v)\bigr)
\label{eq:pbTind}
\eeq
\end{pr}

The proof (see \cite{CISM}) is based on the fact that the factors
$(u-\d_m+S^{(m)})$ have the same Poisson brackets (\ref{eq:pbTind}) which
reproduce
themselves for the product $T(u)$ (Lie-Poisson group structure).

Using the notation \cite{CISM}
$\one T\equiv  T\otimes {\rm id}$, $\two T\equiv  {\rm id}\otimes T$ one can
put the formula (\ref{eq:pbTind}) into a compact form
\beq
 \{\one T(u),\two T(v)\}=\frac{1}{u-v}[{\cal P},\one T(u)\two T(v)]
\label{eq:pbT}
\eeq
where ${\cal P}$ is the permutation operator in
${\bf C}^\N\otimes{\bf C}^\N$.


   Let the spectral invariants $t_\nu(u)$ of the matrix $T(u)$ be defined as
the elementary symmetric polynomials of its eigenvalues
\be
       t_\nu(u)\equiv\tr\bigwedge^\nu T(u), \qquad \nu=1,\ldots,\N
\ee

 For example,
$$ t_1(u)=\tr T(u), \quad t_2(u)=\half(\tr^2T(u)-\tr T^2(u)),\ldots $$

$$      t_N(u)=\det T(u)\equiv d(u).      $$

Note that the central functions $l_\a^{(m)}$ are contained in the determinant
$d(u)\equiv \det T(u)$, see (\ref{eq:specS}).

\begin{pr}
  The non-leading coefficients at powers of $u$ of the polynomials $t_\nu(u)$,
$\nu=1,\ldots,(\N-1)$,
form a commutative, with respect to the Poisson bracket
(\ref{eq:pbT}), family of $M \N(\N-1)/2$ independent Hamiltonians.
\end{pr}

{\bf Proof.}
   The polynomial $t_\nu(u)$
having power $\nu M$ in $u$ contributes $\nu M$ Hamiltonians (its leading
coefficient
is a number), the total number of Hamiltonians is $M(1+2+\cdots+(\N-1))$
$=$ $M\N(\N-1)/2$. 
The commutativity of $t_\nu(u)$
$$      \{t_\mu(u),t_\nu(v)\}=0\qquad     \forall u,v    $$
is a direct consequence of the fundamental Poisson bracket (\ref{eq:pbT}),
see \cite{CISM}. The independence of the integrals of motion is proven in
\cite{RS:1,RS:2} for a different model (Gaudin model, see Section 5) but
the proof is valid also for our case. Note that the assumption made concerning
nondegeneracy of the spectrum of the matrix $Z$ is essential for the
independence of $t_\nu(u)$.

    By virtue of the proposition and since the number of Hamiltonians
constructed
$D=M\N(\N-1)/2$ equals exactly half dimension of the phase space the system is
completely integrable. Now we can turn to the problem of constructing
the separated variables.

\begin{guess} \label{th:AB}
    There exist functions ${\cal A}$ and ${\cal B}$ on $GL(N)$
 such that the following two assertions are true.
First, ${\cal A}(T)$ is an algebraic function and
${\cal B}(T)$, respectively, is a polynomial of degree $D=M\N(\N-1)/2$
of the matrix elements $T_{\a\b}$. Second, the
variables $x_j$, $P_j$ $\quad (j=1,\ldots,D)$ defined from the equations
\beq
 {\cal B}(T(x_j))=0,\qquad  P_j={\cal A}(T(x_j))
\label{eq:defxX}
\eeq
have the Poisson brackets
\beq
   \{x_j,x_k\}=\{P_j,P_k\}=0, \qquad
   \{P_j,x_k\}=P_j\d_{jk}
\label{eq:pbxX}
\eeq
and, besides, are bound to the Hamiltonians $t_\nu(u)$ by the relations
\beq
     \det(P_j-T(x_j))=0
\label{eq:secular}
\eeq
\end{guess}

 The last relation means simply that $P_j$ is an eigenvalue of the
matrix $T(u)$ when $u=x_j$. Putting $P_j=\exp p_j$ we observe that
(\ref{eq:secular}) fits the form (\ref{eq:sepvar})
since the spectral invariants of $T(u)$ contain only  the integrals of motion.

 In the present paper we prove the Conjecture \ref{th:AB}
 for the cases $\N=2$ and $\N=3$.

\section{SL(2) case}
   Though the construction of the separation variables for $\N=2$ is described
in \cite{Skl:14} we reproduce it here in order to fix notation and to prepare
the discussion of more difficult $\N=3$ case.

The system has $M$ degrees of freedom. The spectral invariants of $T(u)$
are
$$ t(u)\equiv t_1(u)\equiv \tr T(u), \qquad
   d(u)\equiv t_2(u)\equiv\det T(u) $$
the trace $t(u)$ containing $M$ integrals of motion.

Define ${\cal A}$ and ${\cal B}$ as \cite{Skl:14}
\beq
 {\cal A}(T)\equiv T_{11} \qquad {\cal B}(T)\equiv T_{12}
\label{eq:defAB}
\eeq
and $x_j$, $P_j$ respectively by the formulas (\ref{eq:defxX}).
For the polynomial $B(u)\equiv{\cal B}(T(u))$ to have $M$ zeroes it is
necessary that its
leading coefficient $Z_{12}$ be nonzero. It can always be done
by a similarity transform $QT(u)Q^{-1}$ which affects neither basic
Poisson brackets (\ref{eq:pbT}), nor Hamiltonians $t(u)$,
since the matrix $Z$, by assumption, has nondegenerate spectrum.

Since the matrix $T(u)$ becomes triangular at $u=x_j$ the quantity $P_j$
is an eigenvalue of $T(x_j)$ and satisfies therefore the secular equation
(\ref{eq:secular}) which in the two-dimensional case takes the form
\be
   P_j^2-t(x_j)P_j+d(x_j)=0 \qquad j=1,\ldots,M
\ee
 Note that the secular
equation defines a {\it hyperelliptic} algebraic curve relating $P_j$
and $x_j$.

To prove the Conjecture \ref{th:AB} it remains to calculate the Poisson
brackets of $P$'s and $x$'s.

\begin{th}
  The Poisson brackets for $P_j$ and $x_j$ are given by (\ref{eq:pbxX}).
\end{th}

   {\bf Proof.}
Let $A(u)={\cal A}(T(u))$ and $B(u)={\cal B}(T(u)).$
Taking particular values of indices in (\ref{eq:pbTind}) one obtains
the identities
\beq
  \{A(u),A(v)\}=0
\label{eq:pbAA}
\eeq
\beq
  \{B(u),B(v)\}=0
\label{eq:pbBB}
\eeq
\beq
  \{A(u),B(v)\}=\frac{A(u)B(v)-B(u)A(v)}{u-v}
\label{eq:pbAB}
\eeq

The commutativity of $B$'s (\ref{eq:pbBB}) entrains obviously the commutativity
of $x_j$ (zeroes of $B(u)$).
The Poisson brackets including $P_j$ can be calculated using implicit
definition of $x_j$. From $B(x_j)=0$ it follows that
$$ 0=\{F,B(x_j)\}=\{F,B(u)\}_{u=x_j}+
                           B^\prime(x_j)\{F,x_j\}  $$
or
$$ \{F,x_j\}=-\frac{\{F,B(u)\}_{u=x_j}}{B^\prime(x_j)}   $$
for any function $F$. In the same way we have
$$ \{P_j,F\}=\{A(x_j),F\}=\{A(u),F\}_{u=x_j}+
                               A^\prime(x_j)\{x_j,F\}      $$
Now it is easy to prove that $ \{P_j,x_k\}=P_j\d_{jk}$.
Starting with
$$ \{P_j,x_k\}=\{A(u),x_k\}_{u=x_j}+A^\prime(x_j)\{x_j,x_k\},$$
expanding the first term further, noting that the second term is
already shown to vanish (\ref{eq:pbBB}), and using (\ref{eq:pbAB}) we arrive at
$$ \{P_j,x_k\}=-\frac{\{A(u),
  B(v)\}_{\scriptstyle u=x_j \atop \scriptstyle v=x_k}}
{B^\prime(x_k)}=
\frac{1}{x_j-x_k}\frac{B(x_j)A(x_k)-A(x_j)B(x_k)}
{B^\prime(x_k)}
$$

The last expression vanishes for $x_j\neq x_k$ due to $B(x_j)=
B(x_k)=0$ and is evaluated via L'H\^opital rule for $x_j=x_k$
to produce the proclaimed result. The commutativity of $P$'s can be shown
in the same way starting from (\ref{eq:pbAA}).

\section{$SL(3)$ case}

Let now $\N=3$. The polynomial $T(u)$ takes values in $3\times3$ matrices.
\be
  T(u)=\left(\begin{array}{ccc}
   T_{11}(u)&T_{12}(u)&T_{13}(u)\\
   T_{21}(u)&T_{22}(u)&T_{23}(u)\\
   T_{31}(u)&T_{32}(u)&T_{33}(u)
\end{array}\right)
\ee

The system has $D=3M$ degrees of freedom.
   The spectral invariants of the matrix $T(u)$
\be
\begin{array}{rclcl}
t_1(u)&\equiv &\tr T(u)&=&\l_1+\l_2+\l_3 \\
t_2(u)&\equiv &\half(\tr^2 T(u)-\tr T^2(u))&=&\l_1\l_2+\l_1\l_3+\l_2\l_3 \\
d(u)&\equiv &\det T(u)&=&\l_1\l_2\l_3
\end{array}
\ee
are the coefficients of the characteristic polynomial for $T(u)$
$$ \det(\l-T(u))=\l^3-t_1(u)\l^2+t_2(u)\l-d(u) $$
which defines a {\it nonhyperelliptic} algebraic curve.

It is convenient to introduce the matrix ${\cal U}(T)$ for any $T\in GL(3)$
\begin{eqnarray*}
  \lefteqn{{\cal U}(T)\equiv T\wedge T \equiv (T^{-1})^t \det T}\\
&&=\left(\begin{array}{ccc}
   T_{22}T_{33}-T_{23}T_{32}&T_{23}T_{31}-T_{21}T_{33}&
             T_{21}T_{32}-T_{22}T_{31}\\
   T_{13}T_{32}-T_{12}T_{33}&T_{11}T_{33}-T_{13}T_{31}&
             T_{12}T_{31}-T_{11}T_{32}\\
   T_{12}T_{23}-T_{13}T_{22}&T_{13}T_{21}-T_{11}T_{23}&
             T_{11}T_{22}-T_{12}T_{21}
   \end{array}\right)
\label{eq:defU}
\end{eqnarray*}
whose elements ${\cal U}_{\a\b}$ are algebraic adjuncts of $T_{\a\b}$.

Let $U(u)={\cal U}(T(u))$.
The Poisson brackets for $T$ and $U$ are calculated easily from (\ref{eq:pbT}):
\beq
 \{\one T(u),\two U(v)\}=-\frac{1}{u-v}[{\cal P}^{t_2},\one T(u)\two U(v)]
\label{eq:pbTU}
\eeq

\begin{eqnarray}
 \lefteqn{\{T_{\a_1\b_1}(u),U_{\a_2\b_2}(v)\}=} \nonumber \\
&&\frac{1}{u-v}\sum_{\gamma=1}^3
   \bigl(-\d_{\a_1\a_2}T_{\gamma\b_1}(u)U_{\gamma\b_2}(v)
+T_{\a_1\gamma}(u)U_{\a_2\gamma}(v)\d_{\b_1\b_2}\bigr)
\label{eq:pbTUind}
\end{eqnarray}

\beq
 \{\one U(u),\two U(v)\}=\frac{1}{u-v}[{\cal P},\one U(u)\two U(v)]
\label{eq:pbUU}
\eeq

\beq
 \{U_{\a_1\b_1}(u),U_{\a_2\b_2}(v)\}=\frac{1}{u-v}
   \bigl(U_{\a_2\b_1}(u)U_{\a_1\b_2}(v)-U_{\a_1\b_2}(u)U_{\a_2\b_1}(v)\bigr)
\label{eq:pbUUind}
\eeq
(the superscript $t_2$ in (\ref{eq:pbTU}) denotes the transposition with
respect to the second space in ${\bf C}^3\otimes{\bf C^3}$).

The experience of the Inverse Spectral Transform Method and, in particular,
$SL(2)$ case suggests that  in $SL(3)$ case
the separated coordinates $x_j$, $j=1,\ldots,3M$ should be defined as zeroes
of some polynomial $B(u)$ of degree $3M$ and the corresponding momenta $p_j$
should be bound to $x_j$ by the secular equation
\be
 P_j^3-t_1(x_j)P_j^2+t_2(x_j)P_j-d(x_j)=0, \qquad P_j=\exp p_j
\ee

It means that $P_j$ should be an eigenvalue of the matrix $T(x_j)$.
Therefore, there must exist such a similarity transformation
\be
T(x_j)\longrightarrow \TT(x_j)=K_jT(x_j)K^{-1}_j
\ee
for each $j$ that the matrix $\TT(x_j)$ is block-triangular
\beq
  \TT_{12}(x_j)=\TT_{13}(x_j)=0
\label{eq:3defx}
\eeq
and $P_j$ is the eigenvalue of $T(x_j)$ splitted from the upper block
\beq
 P_j=\TT_{11}(x_j)
\label{eq:3defX}
\eeq

The problem is reduced thus to determining the polynomial $B(u)$ and the
matrices $K_j$.
Let us take the simplest possible triangular, one-paramet\-ric matrix $K(k)$
\be
 K(k)=\left(\begin{array}{ccc}
            1&k&0\\
            0&1&0\\
            0&0&1
\end{array}\right)
\ee

Note that the matrix
$$
\TT(u,k)\equiv  K(k)T(u)K^{-1}(k)
$$
depends on
{\it two} parameters: $u$ and $k$. Therefore, we can consider the condition
(\ref{eq:3defx}) as the set of two algebraic equations
\beq
  \begin{array}{rclcl}
    \tilde{T}_{12}(x,k)&\equiv& T_{12}(x)+kT_{22}(x)-kT_{11}(x)-k^2T_{21}(x)
       &=&0\\
    \tilde{T}_{13}(x,k)&\equiv& T_{13}(x)+kT_{23}(x)&=&0\\
  \end{array}
\label{eq:eqxk}
\eeq
for two variables $x$ and $k$. Eliminating $k$ from (\ref{eq:eqxk}) one
obtains the polynomial equation for $x$ 
\beq
  T_{23}(T_{12}T_{23}-T_{13}T_{22})-T_{13}(T_{13}T_{21}-T_{11}T_{23})=0
\label{eq:defJ}
\eeq
or
\beq
  T_{23}(x)U_{31}(x)-T_{13}(x)U_{32}(x)=0
\label{eq:eqx}
\eeq

Since the matrix $Z$ is assumed to have simple spectrum,
the leading coefficient ${\cal B}(Z)$ of the polynomial ${\cal B}(T(u))$
can always be made nonzero by a similarity transformation $QT(u)Q^{-1}$,
the equation (\ref{eq:eqx}) being thus of degree $3M$.

Expressing $k$ from $\tilde{T}_{13}=0$ as $k=-T_{13}(x)/T_{23}(x)$
and substituting it into the definition (\ref{eq:3defX}) of $P$ one
obtains
\beq
             P=T_{11}(x)+kT_{21}(x)=-\frac{U_{32}(x)}{T_{23}(x)}
\label{eq:newdefX}
\eeq

So, we have constructed $3M$ pairs of variables $x_j$, $P_j$.
To prove the Conjecture 1 it remains
to show that they have good Poisson brackets.

\begin{th}
  The Poisson brackets for $x_j$ and $P_j$ are given by (\ref{eq:pbxX}).
\end{th}

{\bf Proof.} Let
\beq
 {\cal A}(T)\equiv -\frac{{\cal U}_{32}(T)}{T_{23}} \qquad
   {\cal B}(T)\equiv T_{23}{\cal U}_{31}(T)-T_{13}{\cal U}_{32}(T)
\label{eq:defAB3}
\eeq
Putting $A(u)={\cal A}(T(u))$, $B(u)={\cal B}(T(u))$ and
using (\ref{eq:pbTind}), (\ref{eq:pbTUind}), (\ref{eq:pbUUind})
one can easily calculate the following Poisson brackets:
\beq
   \{A(u),A(v)\}=\{B(u),B(v)\}=0
\label{eq:pbBB3}
\eeq

\beq
   \{A(u),B(v)\}=\frac{1}{u-v}\left(
         A(u)B(v)-B(u)A(v)\frac{T_{23}^2(v)}{T_{23}^2(u)}\right)
\label{eq:pbAB3}
\eeq
from which the wanted Poisson brackets  for $x_j$ and $P_j$
are derived immediately in the same manner as in the $SL(2)$ case.

{\it Remark.} As N. Reshetikhin pointed out to us, the polynomial
${\cal B}(T)$,
see (\ref{eq:defAB3}), is invariant under similarity transform $QTQ^{-1}$
acting on first and second row/column, $Q\in SL(2)\subset SL(3)$. The
$SL(2)$-invariance of ${\cal B}(T)$ entrains invariance of $x_j$ and $P_j$.
The meaning of this fact is still unclear.

\section{Gaudin model}
The above construction of separated variables can be applied also to another
integrable system
--- Gaudin model --- which was introduced first in the quantum variant
\cite{Gaudin}, see also \cite{Skl:24,Jurco:1}. Its classical version turned
out to be a useful example for
developing a general group-theoretic approach to integrable systems
\cite{RS:1,RS:2}.

The model is formulated in terms of the same $SL(\N)$ variables
$S_{\a\b}^{(m)}$ as in Section 2, see (\ref{eq:pbS}). Consider the matrix
function
\be
  \T(u)\equiv {\cal Z}+ \sum_{m=1}^M \frac{S^{(m)}}{u-\d_m}
\ee
where $\{\d_m\}_{m=1}^M$ are some fixed parameters and ${\cal Z}$ is a
traceless number matrix having $\N$ distinct eigenvalues.
 In contrast with $T(u)$, see (\ref{eq:pbT}), the matrix
$\T(u)$ has linear Poisson brackets.

\begin{pr} \label{th:pbTT}
\beq
\label{eq:pbTT}
 \{\one \T(u),\two \T(v)\}=\frac{1}{u-v}[{\cal P},\one \T(u)+ \two \T(v)]
\eeq
\end{pr}

The proof is a matter of direct computation \cite{CISM}.

Consider now the spectral invariants $\tau_\nu(u)$ of the matrix $\T(u)$
\be
  \tau_\nu(u)\equiv \tr\T(u)^\nu, \qquad \nu=1,\ldots,\N
\ee

Note that $\tau_\nu(u)$ is a meromorphic function of $u$
\be
   \tau_\nu(u)=\zeta_\nu+\sum_{m=1}^M\sum_{\a=1}^\nu
                    \frac{\tau_{m,\nu}^\a}{(u-\d_m)^\a}
\ee

Note that $\zeta_\nu=\tr{\cal Z}^\nu$ and $\tau_{m,\nu}^\nu=\tr[S^{(m)}]^\nu$
$=\sum_{\a=1}^\N [l_\a^{(m)}]^\nu$ are numbers, see (\ref{eq:specS}).
The following proposition, analogous to Proposition 2, states the complete
integrability of the system.

\begin{pr} \label{th:intTT}
  The quantities $\tau_{m,\nu}^\a$, $(m=1,\ldots,M$; $\nu=2,\ldots,\N$;
$\a=1,\ldots,(\nu-1))$, form a commutative, with respect to the Poisson bracket
(\ref{eq:pbTT}), family of $M\N(\N-1)/2$ independent Hamiltonians.
\end{pr}

{\bf Proof.} It is easy to compute the total number of Hamiltonians
$M(1+2+\cdots+(\N-1))=M\N(\N-1)/2$ which equals exactly half dimension of the
phase space. The commutativity of the spectral invariants of $\T(u)$
follows directly from (\ref{eq:pbTT}), see \cite{RS:1,RS:2}. The independence
of the Hamiltonians  is also proven there.

The analog of Conjecture 1 for the Gaudin model is presented below.

\begin{guess}
Let ${\cal A}$ and ${\cal B}$ be the same functions on $GL(N)$ that  in
the Conjecture 1. Then the variables $x_j$ and $p_j$ defined by the equations
\beq
   {\cal B}(\T(x_j))=0, \qquad p_j={\cal A}(\T(x_j))
\label{eq:defxiXi}
\eeq
have the canonical Poisson brackets (\ref{eq:px})
and, besides, are bound to the Hamiltonians $\tau_{m,\nu}^\a$ by the
relation  $\det(p_j-\T(x_j))=0$.
\end{guess}

The separation of variables for $\N=2$ and $\N=3$ cases is performed now in the
 same manner as, respectively, in Sections 3 and 4.

\begin{th}
The Conjecture 2 is true for $N=2$.
\end{th}

  {\bf Proof.}   In the $SL(2)$ case,
the functions ${\cal A}$ and ${\cal B}$ on $GL(2)$ are defined by the formulas
(\ref{eq:defAB}).
Like in Section 3, we can always suppose that ${\cal B}({\cal Z})\neq0$.
The variables $x_j$, $p_j$, $j=1,\ldots,\N$ are then determined  by the
equations (\ref{eq:defxiXi}).

Let $A_G(u)={\cal A}(\T(u))$ and $B_G(u)={\cal B}(\T(u))$.
Taking particular matrix elements of (\ref{eq:pbTT}) one
obtains
\beq
  \{{A}_G(u),{A}_G(v)\}=
  \{{B}_G(u),{B}_G(v)\}=0
\eeq

\beq
  \{{A}_G(u),{B}_G(v)\}=-\frac{{B}_G(u)-{B}_G(v)}{u-v}
\eeq

  The rest of the proof follows that of the Theorem 1.

\begin{th}
The Conjecture 2 is true for $N=3$.
\end{th}

 {\bf Proof.}   In the $SL(3)$ case define  ${\cal A}$, ${\cal B}$
by the formulas (\ref{eq:defAB3}) and, like in Section 4,
 suppose that ${\cal B}({\cal Z})\neq0$.

Let again $A_G(u)={\cal A}(\T(u))$ and $B_G(u)={\cal B}(\T(u))$.
It suffices to establish the Poisson brackets
\beq
  \{A_G(u),A_G(v)\}=
  \{B_G(u),B_G(v)\}=0
\label{eq:pbAABBg}
\eeq

\beq
   \{A_G(u),B_G(v)\}=\frac{1}{u-v}\left(
         B_G(v)-B_G(u)\frac{T_{23}^2(v)}{T_{23}^2(u)}\right)
\label{eq:pbABg}
\eeq
since the remaining calculation is standard.

   It is possible to verify the above Poisson brackets directly, using
(\ref{eq:pbTT}).
It is simpler, however, to avoid long computations and to use the fact
\cite{Gaudin,Skl:24}
 that the Gaudin model is in fact a degenerate case of the magnetic chain.
To be precise, let us replace $S^{(m)}$ in (\ref{eq:defT}) by $\eps S^{(m)}$,
$Z$ by $1+\eps{\cal Z}$,
and divide $T(u)$ by $\prod_{m=1}^M(u-\d_m)$. Then, in the first order in
$\eps$, we have
$T(u)/\prod_{m=1}^M(u-\d_m)$ $= 1+\eps \T(u)+O(\eps^2)$.
The Poisson brackets (\ref{eq:pbTT}) are obtained, respectively, as the
linearization of the quadratic Poisson brackets (\ref{eq:pbT}).

To conclude the proof, it remains to notice that
$$ A(u)=1+\eps A_G(u)+O(\eps^2) \qquad
   B(u)=\eps^3 B_G(u)+O(\eps^4)
$$
and that the Poisson brackets (\ref{eq:pbAABBg}), (\ref{eq:pbABg}) are
obtained in the
leading order in $\eps$ from (\ref{eq:pbBB3}), (\ref{eq:pbAB3}).

\section{Unsolved problems}

   The natural question arises whether the construction of separated
variables presented here for the $SL(2)$ and $SL(3)$ cases can be generalized
to the $SL(\N)$ case and further, to the integrable systems associated with
classical $r$-matrices corresponding to other simple Lie algebras.
Hopefully, the generalization will elucidate the geometric and algebraic
meaning of the construction.

The
$SL(\N)$ case is presently under study. The problem consists in finding a
 multiparametric family of matrices $K(k_1,\ldots,k_Q)$ such that
after eliminating $k$'s from the system $T_{12}(x)=\ldots=T_{1\N}(x)=0$
the resulting equation for $x$ provide the necessary number of commuting
zeroes. Another challenging object of study is the Kowalewski top which can be
considered as a Gaudin model for $Sp(4)\simeq SO(5)$ group \cite{Kowal}.

   Since the construction of separated variables for the $SL(2)$ case
has the direct quantum counterpart \cite{Skl:30,Skl:32,Skl:14}, it seems
reasonable to conjecture that the same is true for the $SL(\N)$ case.

\bigskip
\noindent{\it Acknowledgments.}
I am grateful to A. Bobenko, J. Hietarinta, V. Kuzne\-tsov, N. Reshetikhin and
A. Reyman for valuable and encourageing discussions.


\end{document}